\documentclass{moriond}
\usepackage{amsmath,bm}

\def\be{\begin{equation}}
\def\ee{\end{equation}}
\def\bea{\begin{eqnarray}}
\def\eea{\end{eqnarray}}

\newcommand{\Photo}{}

\begin{document}
\vspace*{4cm}
\title{{\em Ab initio} modeling of Galactic dust polarized CMB foreground}

\author{Alexei G. Kritsuk$^1$, Ka Wai Ho$^2$, Ka Ho Yuen$^3$ \footnote{Present address: Nanjing University, Suzhou, Jiangsu Province, People's Republic of China}, and Raphael Flauger$^1$}
\address{$^1$Physics Department, University of California, San Diego, La Jolla, CA 92093, USA\\
$^2$Kavli Institute for Theoretical Physics, University of California, Santa Barbara, CA 93106, USA\\
$^3$Theoretical Division, Los Alamos National Laboratory, Los Alamos, NM 87545, USA}

\maketitle

\abstracts{We present the analysis of high-resolution synthetic dust polarization maps derived from large-scale simulations of magnetized multiphase interstellar turbulence carried out with the \textsf{AthenaK} code on the {\em Frontier} exascale supercomputer at the Oak Ridge National Laboratory. Our turbulence model accurately captures spectral properties of the $E$- and $B$-modes measured by {\em Planck} at 353~GHz. The simulations provide new insights into the physical origins of the observed $E/B$ asymmetry and positive $TE$ signal, facilitating the development of advanced models of Galactic foreground emission for current and future CMB experiments.}

\section{Introduction}
Polarized emission from Galactic dust is one of the dominant foregrounds for cosmic microwave background (CMB) polarization and a major challenge for extracting the primordial CMB signal on large angular scales. 

In this short communication we report results from a high-resolution ($4096^3$) numerical simulation of turbulence in the multiphase interstellar medium (ISM). Our approach employs a homogeneous multiphase MHD turbulence model~\cite{kritsuk..17}, which successfully reproduced a number of statistics for the local ISM, to generate synthetic dust polarization maps~\cite{kritsuk..18} of the $I$, $Q$, and $U$ Stokes parameters. The high-fidelity maps are then used to decompose the $E$ and $B$ modes and compute the $TT$, $EE$, $BB$, $TE$, and $TB$ spectra, $EE/BB$ and $TE/EE$ spectral ratios, and $r^{TE}\left(\equiv C^{TE}(k)/\sqrt{C^{TT}(k)C^{EE}(k)}\right)$, $r^{TB}$ correlation ratios. We then compare these spectral characteristics of the polarized foreground dust emission to {\em Planck} PR3 measurements at 353~GHz~\cite{planckXI-20} for the LR71 region, covering 71\% of the sky, and find an excellent match for all of them, except $r^{TB}$. This comparison, in turn, allows us to resolve several outstanding puzzles concerning the physical origins of the observed $E/B$ asymmetry and positive $TE$ correlation with an aid of two complementary simulations of highly supersonic, trans- and sub-Alfv\'enic MHD turbulence with an isothermal equation of state.

This work further solidifies conclusions based on our previous lower resolution simulations~\cite{kritsuk..18,ho...25} by achieving more realistic effective Reynolds numbers, approaching the characteristic values for the local ISM at the solar radius in the Milky Way disk.

\section{Numerical model}
\begin{figure}
\centerline{\includegraphics[width=0.63\linewidth]{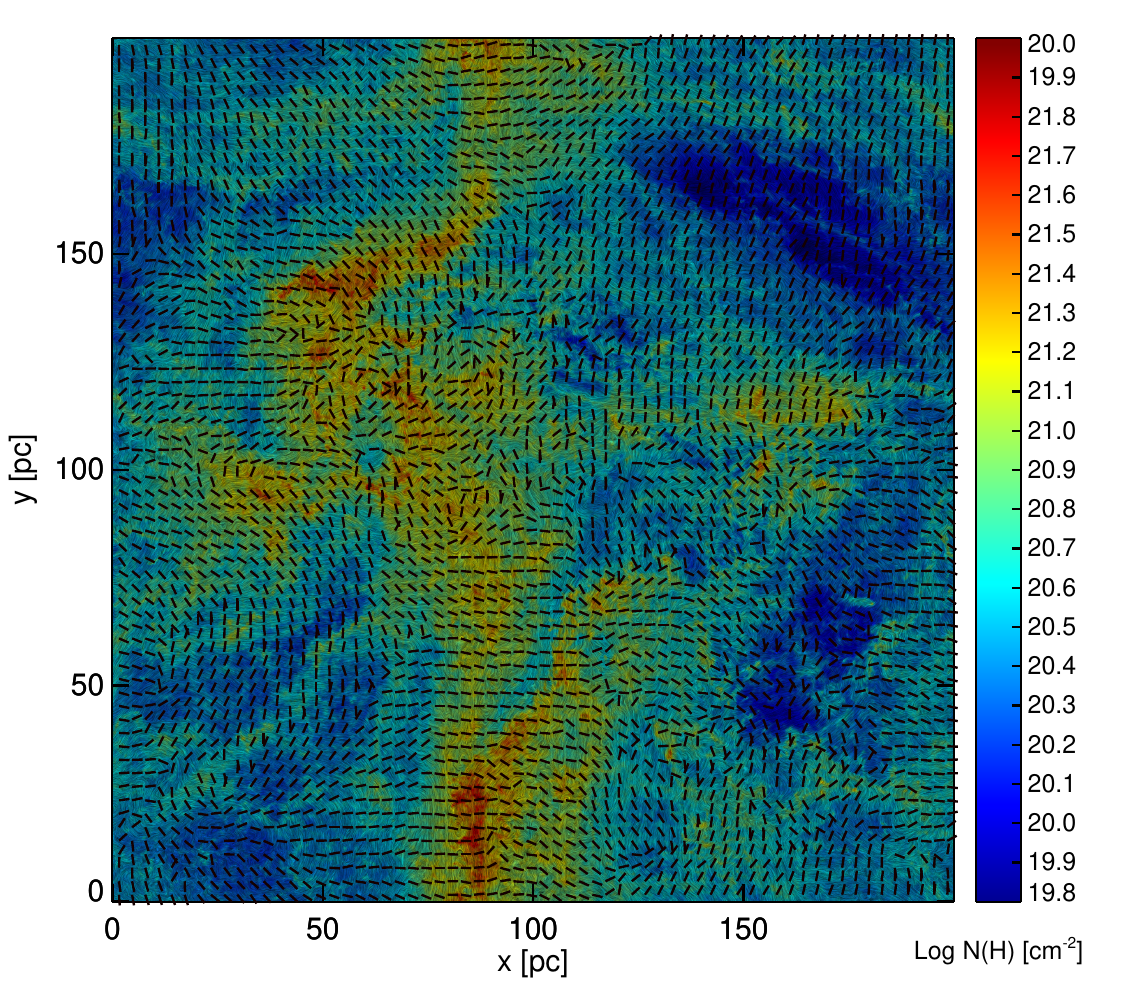}}
\caption[]{Sample synthetic map showing the plane-of-sky (POS) magnetic field structure (drapery texture) and H{\sc i} column density at $t=37$~Myr for the projection along the mean field $\bm b_0=6\;\mu$G. Pseudovectors indicate the polarization direction  (predominantly perpendicular to the POS field). The map is built on reduced-resolution $2048^3$ numerical data additionally smoothed with a low-pass box-car filter of length 5 voxels.}
\label{map}
\end{figure}

Our ISM turbulence model depends on four parameters, which are largely constrained by the observations, namely the size of periodic domain, $L=200$~pc, the 3D velocity dispersion at this scale, $u_{\rm rms}=20$~km/s~\cite{koley23}, the mean magnetic field strength at this scale, $b_0=6$~$\mu$G, and the mean H{\sc i} number density, $\langle n_{\rm HI}\rangle=1$~cm$^{-3}$. The choice of $L$ is motivated by the scale height of the cold neutral medium, $\sim100$~pc~\cite{dickey22,smith.......23}, which is close to the energy injection scale associated with active supernova feedback. To make the model tractable, we adopt a simplified approach and apply a random external large-scale solenoidal forcing to induce energy cascade. An artificial kinetic energy injection algorithm based on the Ornstein-Uhlenbeck stochastic process is known to stir up a specific form of spatio-temporal intermittency in MHD turbulence, leading to the development of fat stretched-exponential tails in the probability density function (PDF) of magnetic field fluctuations~\cite{kritsuk..17}. To get rid of the spurious effects of this intermittency in synthetic polarization maps~\cite{ho...25}, we mask out the extreme end of high density voxels ($n_{\rm HI}>n_{\rm t}$, where $n_{\rm t}=70$~cm$^{-3}$ is the threshold H{\sc i} number density of the mask).

The $4096^3$ multiphase simulation was carried out with a version of the \textsf{AthenaK} code~\cite{stone...20}, using the {\tt PPM-4} reconstruction, {\tt HLLD} Riemann solver, {\tt IMEX2} time integration, and constrained transport (CT)~\cite{gardiner.05}. Our production runs covered 75~Myr of evolution from static, uniform initial conditions and utilized 4096 nodes (32k GPU equivalents) of the {\em Frontier} system at ORNL. In total 228 sets of synthetic maps were generated on the fly for statistically stationary fully developed MHD turbulence (uniformly sampled in time for $t\in[30,75]$~Myr; the large eddy turnover time $\tau_{\rm d}=5$~Myr and the forcing auto-correlation time $\tau_{\rm a}=2$~Myr).
The two archival $1024^3$ isothermal simulations~\cite{kritsuk...09a,kritsuk...09} were carried out with the {\tt PPML} solver~\cite{ustyugov...09,kritsuk11}, which also employed HLLD and CT, and covered about $6\tau_d$ of stationary evolution.

\section{Results}
A sample synthetic polarization map taken at $t=38$~Myr is shown in Fig.~\ref{map}. As expected, the thermal dust polarization direction is mostly perpendicular to the direction of plane-of-sky (POS) magnetic field shown by the drapery pattern. Likewise, the POS field direction is predominantly parallel to the dense filaments in this map, as expected for lower-density, diffuse regions of the ISM, where gravity does not play a significant role.

\begin{figure}
\begin{minipage}{0.50\linewidth}
\centerline{\includegraphics[width=1.0\linewidth]{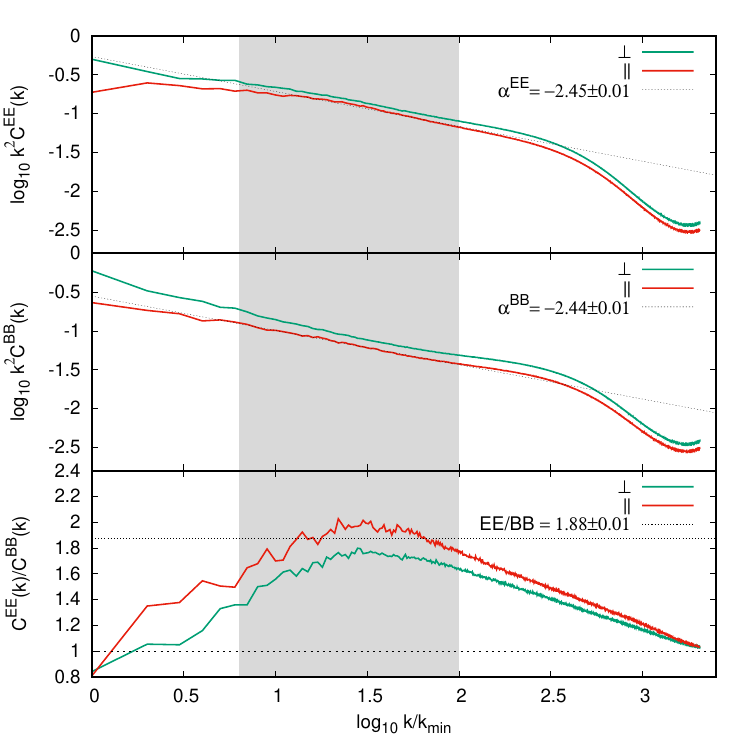}}
\end{minipage}
\hfill
\begin{minipage}{0.445\linewidth}
\centerline{\includegraphics[width=1.0\linewidth]{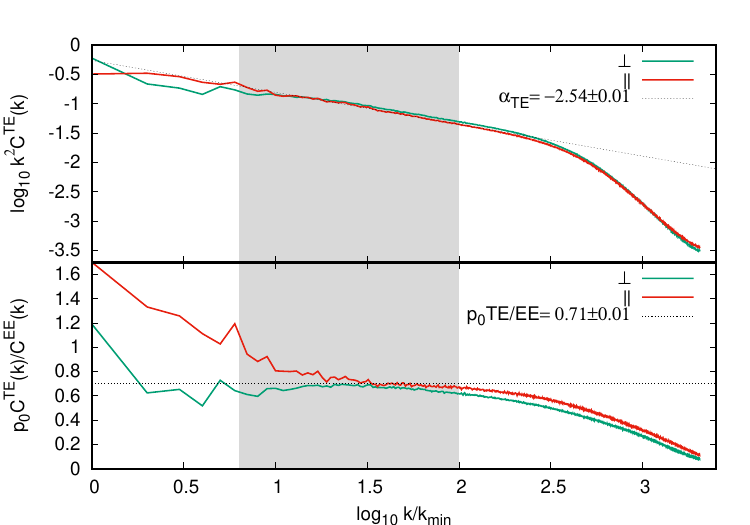}}
\centerline{\includegraphics[width=1.0\linewidth]{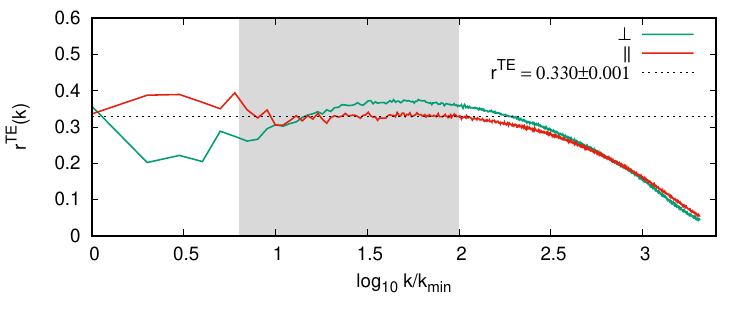}}
\end{minipage}
\caption[]{Compensated $EE$ and $BB$ spectra and $EE/BB$ power ratios (left) for lines of sight parallel ($\parallel$, red) and perpendicular ($\perp$, green) to the direction of the mean magnetic field $\bm b_0$. Same for the $TE$ spectra and $TE/EE$ power ratios (top right) and for the correlation ratio $r^{TE}(k)$ (bottom right). The range of wave numbers $\log_{10}(k/k_{\rm min})\in[0.8,2.0]$ shown with a shade of grey corresponds to the inertial range with sub- or trans-Alfv\'enic turbulence, reflecting the conditions expected in the local ISM. Dotted lines show least-squares fits to the $\parallel$ spectra and power ratios obtained within the shaded range of $k$.}
\label{teb}
\end{figure}

Figure~\ref{teb} shows time-averaged spectra and spectral ratios derived from the synthetic maps. Spectral slopes $\alpha_{EE}=-2.45$,  $\alpha_{BB}=-2.44$, and  $\alpha_{TE}=-2.54$  obtained in the range of scales where approximate equipartition of magnetic and kinetic energy is reproduced (shaded band in the figure) resemble {\em Planck} PR3 scaling best measured at 353~GHz in the largest sky region LR71~\cite{planckXI-20}: $-2.42\pm0.02$, $-2.54\pm0.02$, and $-2.50\pm0.02$, respectively.

\begin{figure}
\begin{minipage}{0.45\linewidth}
\centerline{\includegraphics[width=1.0\linewidth]{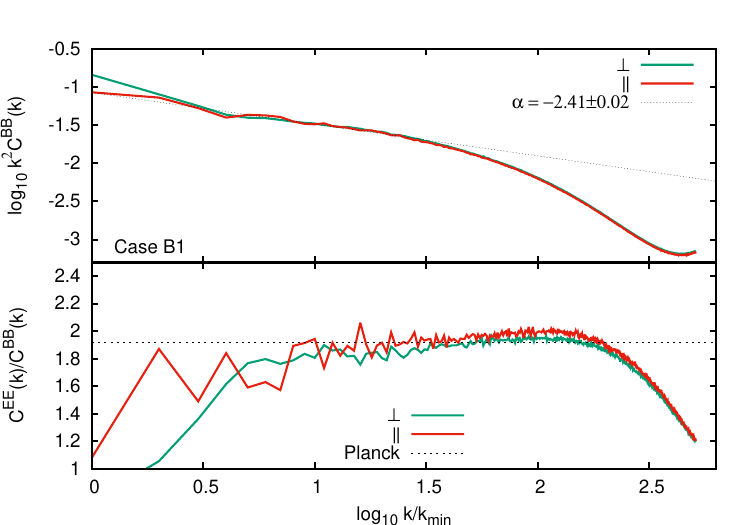}}
\end{minipage}
\hfill
\begin{minipage}{0.45\linewidth}
\centerline{\includegraphics[width=1.0\linewidth]{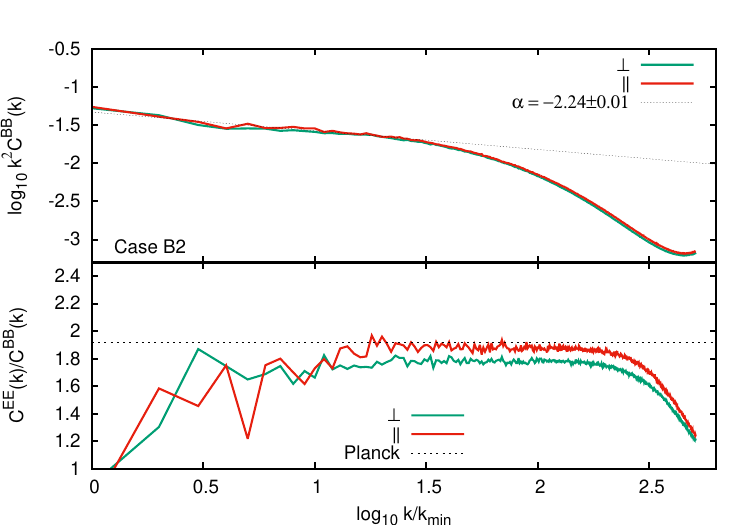}}
\end{minipage}
\caption[]{$B$-mode spectra and $EE/BB$ power ratios for isothermal cases B1 (left; $M_{\rm s}=10$, $M_{\rm a}=0.6$) and  B2 (right; $M_{\rm s}=9$, $M_{\rm a}=1.0$), both with a mask threshold $n_{\rm t}/n_0=20$ leading to $\sim0.2$\% voxels masked.
Without masking, $EE/BB\approx1.6$ and 1.7 for cases B1 and B2, respectively.}
\label{i-teb}
\end{figure}

The $EE/BB$ power ratio measured in the same interval of scales, $EE/BB=1.88$, is very close to {\em Planck's} $EE/BB=1.89\pm0.03$, although in the simulation there are clear hints of scale-dependence with a tendency of decreasing $EE/BB$ ratios toward smaller scales. While higher-resolution multiphase simulations are certainly needed to confirm the trend, we note that our isothermal simulations (Fig.~\ref{i-teb}) in both sub- and trans-Alfv\'enic regimes do not show the same scale-dependence in the far-inertial range. We argue that, keeping in mind the `shocking' origin of the $E/B$ asymmetry~\cite{flauger..26}, one can explain the difference as due to substantially lower sonic Mach numbers, $M_{\rm s}$, in a more realistic multiphase model ($M_{\rm s}\approx4$) compared to the  strongly supersonic isothermal cases B1 ($M_{\rm s}\approx10$) and B2 ($M_{\rm s}\approx9$), in which the sonic scale cannot be resolved on a $1024^3$ grid. 

In contrast, the $TE/EE$ and $r^{TE}$ ratios are both approximately constant across the inertial range in the multiphase simulation (Fig.~\ref{teb}, right panels). Comparing our $TE/EE=0.71/p_0$ with {\em Planck's} measurement $TE/EE=2.77\pm0.05$, we estimate the value of intrinsic polarization, $p_0= 0.26$. Note that our model prediction for the $TE$ correlation ratio, $r^{TE}=0.33$, is pretty close to {\em Planck's} result $r^{TE}=0.357\pm0.003$. Finally, we find $r^{TB}=-0.0015\pm0.0006$, which is consistent with the parity-preserving design of our turbulence model and predictably inconsistent with {\em Planck's} measurement $r^{TB}\approx0.05$.

\section{Conclusions and perspective}
In this short communication we presented first results from our new high-resolution ($4096^3$) simulation of multiphase ISM turbulence, which generated a large array of synthetic dust polarization maps. We used the maps to derive spectral properties for the $E$- and $B$-modes and show that our model accurately captures {\em Planck} results for polarized dust foregrounds at 353~GHz for the LR71 sky region\,\cite{planckXI-20}, with a single exception of the finite $TB$ correlation ratio $r^{TB}\approx0.05$ that would require a parity-violating feature currently lacking in the  model. Our model suggests a decreasing $EE/BB$ ratio toward the far-inertial range in ISM turbulence, while the $TE$ correlation remains approximately constant across the scales. More realistic higher resolution simulations are needed to accurately capture the observed $r^{TE}$ correlation ratio and polarization fraction PDF as well as scale dependence of the $E/B$ mode asymmetry.

\section*{Acknowledgments}
This research was supported in part by the NASA Grant No. 80NSSC22K0724 (AK, RF) and by grants 216179 from the Simons Foundation and NSF PHY-2309135 to the Kavli Institute for Theoretical Physics (KITP).
Computational and storage resources were provided by the INCITE-2025 and DD awards at ORNL (ast193), by the LRAC award at TACC (AST21004), by the DOE award allocated at NERSC (FES-ERCAP-m4239), 
and by the ACCESS resource allocation at SDSC and TACC (MCA07S014).

\section*{References}
\bibliography{kritsuk}


\end{document}